\documentclass{elsart}
\usepackage{graphicx}
\usepackage{amsmath}
\usepackage{amssymb}

\begin{document}
\begin{frontmatter}
\title{Correlation energy, pair-distribution functions and static 
structure factors of jellium}

\author{Paola Gori-Giorgi}
and
\author{Francesco Sacchetti}

\address{Dipartimento di Fisica and Unit\`a INFM, Universit\`a di Perugia,
         Via A. Pascoli 1, 06123 Perugia, Italy}

\author{Giovanni B. Bachelet\thanksref{mail1}}
\address{Dipartimento di Fisica and Unit\`a INFM,
         Universit\`a di Roma ``La Sapienza'', Piazzale Aldo Moro 2, 00185
         Rome, Italy}

\thanks[mail1]{corresponding author, {\tt Giovanni.Bachelet@roma1.infn.it}}

\begin{abstract}
We discuss and clarify a simple
and accurate interpolation scheme for the spin--resolved electron
static structure factor (and corresponding pair correlation
function) of the 3D unpolarized homogeneous electron gas which,
along with some analytic properties of the spin--resolved pair--correlation
functions, we have just published~\cite{noi}. We compare 
our results with the very recent
spin--resolved scheme by Schmidt 
{\sl et al.}~\cite{PerdewWang,comment}, 
and focus our attention on the 
spin--resolved correlation energies
and the high--density limit of the correlation functions. 
\end{abstract}


\end{frontmatter}

\section{Introduction}
The homogeneous electron gas is a model solid whose positive ionic
charges are smeared throughout the whole crystal volume to yield a
shapeless, uniform positive background (whence the nickname of
jellium). The
model, by ignoring the ionic lattice which makes real materials
different from one another, allows the theorists to concentrate on key
aspects of the electron--electron interaction. It thus
represents an obvious limit for the inhomogeneous
electron gas, and, through the Density Functional
Theory (DFT)~\cite{DFTbook}, its Local Density Approximation (LDA)
and other semi--local~\cite{GGA} and nonlocal~\cite{Gunn}
approximations, it links to a
popular and very successful description of real materials.

The pair--distribution functions $g_{\sigma_1
\sigma_2}({\bf r}_1,{\bf r}_2)$ describe  the spatial correlations 
between electron pairs of prescribed spin orientations: the expected 
number of spin--$\sigma_2$ electrons
in the volume $dV$ at ${\bf r}_2$, when another
electron of spin $\sigma_1$ is at ${\bf r}_1$, is
equal to $dN({\bf r}_2\sigma_2|{\bf r}_1\sigma_1)=
n_{\sigma_2}({\bf r}_2)\,g_{\sigma_1\sigma_2}({\bf r}_1,{\bf r}_2)\,dV$,
where $n_{\sigma}({\bf r})$ is the
density of spin--$\sigma$ electrons. 
In the spin--unpolarized jellium, the electronic spin density
$n_{\uparrow}({\bf r})= n_{\downarrow}({\bf r})=n/2=(8\pi
r_s^3/3)^{-1}$ is uniform in space (i.e.  independent of ${\bf r}$),
so $g_{\sigma_1 \sigma_2}({\bf r}_1,{\bf
r}_2)$ only depends on the distance between the two electrons $r=|{\bf
r}_1-{\bf r}_2|$. Hartree atomic units are used throughout this work.
The static structure factor $S(q)$ is an ``experimental''
quantity which gives a measure
of the instantaneous density correlations in the system, and is directly
related to the Fourier transform of the pair--correlation function.
For an unpolarized homogeneous electron gas, after introducing the
Fermi wavevector $q_F=(3\pi^2\,n)^{1/3}$, the scaled variables 
$\rho=q_F\,r$ and $k=q/q_F$ are
often convenient.  With these variables the 
spin--resolved static structure factors are
written as
\begin{equation}
S_{\sigma_1 \sigma_2}(k;r_s)  =  \delta_{\sigma_1,\sigma_2}
+\frac{2}{3\,\pi}\int_0^\infty d\rho\,\rho^2\,
\left[g_{\sigma_1 \sigma_2}(\rho;r_s)-1\right]\,\frac{\sin (k\,\rho)}
{k\,\rho};\label{S_def}
\end{equation}
the total pair--distribution function is equal to
$g=\frac{1}{2}(g_{\uparrow\uparrow}+g_{\uparrow\downarrow})$,
and the total static structure factor to $S=S_{\uparrow\uparrow}+
S_{\uparrow\downarrow}$.\par
The pair--distribution functions of the uniform electron gas
are a key ingredient in the construction of semi--local
and nonlocal density functionals~\cite{GGA,Gunn}.
At the densities of interest for DFT calculations, the best estimate
for the pair--correlation functions and static
structure factors\footnote{The pair--correlation functions
and static structure factors are independently extracted
by quantum Monte Carlo simulations~\cite{OB}.} of jellium is given by
quantum Monte Carlo simulations (QMC)~\cite{OB,OBH} which are available
for a discrete set of interelectronic distances $\rho$ 
(or momentum transfer $k$) and
densities $r_s$. In a recent work~\cite{noi} we have presented
simple functional forms for the spin--resolved pair--correlation
functions which depend upon $\rho$ and $r_s$, are
analytic and closed--form both in real and reciprocal space, fulfill
most of the known properties of their exact counterparts, and contain
some free paramaters that have been fixed by a two--dimensional ($\rho,
r_s$ and $k,r_s$) 
fit to the new QMC data~\cite{OBH}, thus yielding
very accurate and reliable functions in the relevant density
range $0.1\lesssim r_s \le 10$. As a byproduct, we also obtained 
accurate spin--resolved correlation energies which fulfill 
the exact high--density
limit by construction~\cite{noi}.
In Sec.~\ref{sec_prop}
we summarize the small--$\rho$ ($k$)
behavior of $g_{\sigma_1\sigma_2}$ ($S_{\sigma_1\sigma_2}$)
together with the corresponding large--$k$ ($\rho$) behavior of 
$S_{\sigma_1\sigma_2}$ ($g_{\sigma_1\sigma_2}$), and we discuss
and clarify some points, especially Eq.~(12) of Ref.~\cite{noi}. 
Sec.~\ref{sec_PWnew} is devoted
to a brief comparison of our results with the new 
spin--resolution~\cite{comment} of the Perdew--Wang
function~\cite{PerdewWang}, and in Sec.~\ref{sec_energy}
the corresponding spin--resolved correlation energies are 
discussed.

\section{Behavior of the spin--resolved correlation functions 
for small and large arguments} 
\label{sec_prop}
The pair--distribution function $g=g_{\rm ex}+g_c$ 
(and correspondingly the static structure factor) can be divided into an
exchange--only contribution $g_{\rm ex}(\rho)$ 
(given by the Hartree--Fock approximation) and a Coulomb--correlation
contribution $g_c(\rho;r_s)$, which, in turn, can be split into
its $\uparrow\downarrow$ and $\uparrow\uparrow$ parts, 
$g_c=\frac{1}{2}(g_{\uparrow\downarrow}^c+g_{\uparrow\uparrow}^c)$.
The leading terms of $g^c_{\sigma_1\sigma_2}$,
$S^c_{\sigma_1\sigma_2}$, $g_{\rm ex}$,
$S_{\rm ex}$ and of the total functions 
for small and large arguments are 
summarized below. 
\[
\hspace{-0.3cm}\begin{array}{l|c|c|c|c}
& g(\rho \to 0) & S(k \to \infty) &   
g(\rho \to \infty) & S(k\to 0) \\
\hline
{\rm corr.}\uparrow\downarrow & a\,\Big(1+\dfrac{\rho}{q_F}\Big)-1
& -\dfrac{4}{3\pi q_F}\dfrac{a}{k^4} &   
\dfrac{9}{4}\Big(\dfrac{1}{\rho^4}+\dfrac{1}{\rho^6}\Big)
& -\dfrac{3}{8}\,k+
\dfrac{k^2q_F^2}{4\omega_p}+\dfrac{k^3}{32} \\
{\rm corr.}\uparrow\uparrow & b
\Big(\rho^2+\dfrac{\rho^2}{2q_F}\Big)
-\dfrac{\rho^2}{5}
& \dfrac{8}{\pi q_F}\dfrac{b}{k^6} & 
\dfrac{9}{4}\Big(\dfrac{1}{\rho^4}+\dfrac{1}{\rho^6}\Big)
& -\dfrac{3}{8}k+
\dfrac{k^2q_F^2}{4\omega_p}+\dfrac{k^3}{32}\\
{\rm exch.} & \dfrac{1}{2}+\dfrac{1}{10}\rho^2 & 1 
& 1-\dfrac{9}{4}\Big(\dfrac{1}{\rho^4}+\dfrac{1}{\rho^6}\Big)
& \dfrac{3}{4}k-\dfrac{k^3}{16} \\
{\rm total} & \dfrac{a}{2}\Big(1+\dfrac{\rho}{q_F}\Big)
& 1-\dfrac{4}{3\pi q_F}\dfrac{a}{k^4} & 1-3\,d\,\dfrac{6!}{\rho^8}
&   \dfrac{k^2q_F^2}{2\omega_p}+c\,k^4+d\,k^5
\end{array}
\]
Here $\omega_p=\sqrt{3/r_s^3}$ is the classical plasma frequency and
the constants $a$, $b$, $c$ and $d$ are not known.
The small--$\rho$ (and corresponding large--$k$) behavior 
of the correlation functions
is well known~\cite{Kimball2} from
the many--body Schr\"odinger equation when two electrons
approach each other (cusp conditions).
 The small--$k$ (and corresponding 
large--$\rho$) behavior of the $\uparrow\downarrow$ and
$\uparrow\uparrow$ correlation functions
seems, instead, to be less known~\cite{noi}. It can be determined by means
of the random phase approximation (RPA) (see e.g. Ref.~\cite{Pines}), which
is exact in the $k\to 0$ limit,
as follows. The RPA only takes into account direct processes, i.e.
processes that occur for both parallel-- and antiparallel--spin
pairs. Thus, for an unpolarized gas, the RPA 
$\uparrow\downarrow$ and $\uparrow\uparrow$ correlation functions
are equal. This simple consideration, together with
the well--known small--$k$ behavior of the total
$S$, $q_F^2k^2/2\omega_p$, tells us that, as $k\to 0$,
the linear term $3k/4$ of $S_{\rm ex}$ must be canceled
50\% by $S^c_{\uparrow\downarrow}$ and 50\% by
$S^c_{\uparrow\uparrow}$. The same argument can be
applied to the $-k^3/16$ term of $S_{\rm ex}$; in fact, 
 no term $\propto k^3$ must 
appear in the total $S$, whose long--wavelength
behavior is determined by the plasmon contribution and by
the single--pair and multipair quasiparticle--quasihole excitation
contributions, proportional to $k^5$ and $k^4$
respectively~\cite{Pines,iwa}. While
the leading $-3k/8$ term of
$S^c_{\sigma_1\sigma_2}$ (which corresponds to the
large--$\rho$ term $\frac{9}{4}\rho^{-4}$) and the plasmon
contribution $\propto k^2$ must hold beyond RPA, the $k^3/32$
term (which corresponds to the large--$\rho$ term $\frac{9}{4}
\rho^{-6}$) holds for the exact $S^c_{\sigma_1\sigma_2}$ 
in the high density limit, but its validity at lower densities
must be verified. Notice also that the $k^5$ term in the
$k\to 0$ expansion of the total static structure factor implies
that the total $g$ behaves like $\rho^{-8}$ when $\rho\to\infty$.
The functional forms proposed in 
Ref.~\cite{noi} exactly fulfill all of the above analytic constraints.

\section{Spin--resolved correlation functions}
\label{sec_PWnew}
\begin{figure}[b]
\begin{center}
\hspace{-0.55cm}
\includegraphics[scale=0.8]{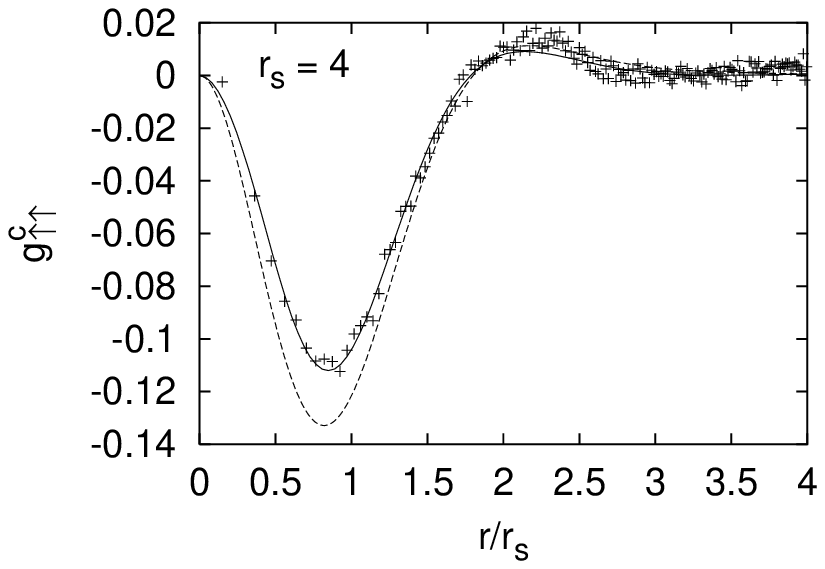}
\includegraphics[scale=0.8]{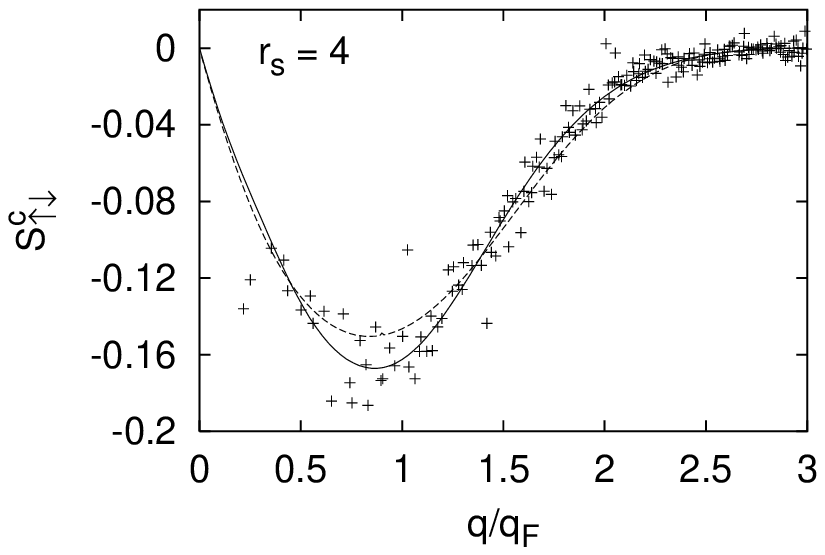}
\end{center}
\caption
{Parallel--spin contribution to the pair--correlation function (left), and
antiparallel--spin contribution to the static structure factor (right) for
$r_s=4$. Solid line: functions from Ref.~\protect\cite{noi}; dashed line: new 
spin--resolved PW model~\protect\cite{PerdewWang,comment};
crosses: QMC data~\protect\cite{OBH}.}
\label{fig_newPW}
\end{figure}

In Ref.~\cite{noi} we compared our correlation functions
with the widely used Perdew--Wang (PW)
\cite{PerdewWang} model. The PW function turned out not to be accurate in its
spin--resolved version, mainly because it does not fulfill the
exact $k\to 0$ limit of $S^c_{\sigma_1\sigma_2}$. Based on our work,
Schmidt {\sl et al.}~\cite{comment} have recently proposed
a new spin resolution of the PW model, obtained by imposing
this exact limit. Such revised PW function represents
a considerable improvement 
over the original one, and works very well for
$r_s\lesssim 2$, both in real and reciprocal space. At lower densities,
however, our functions~\cite{noi} provide a better interpolation of the QMC
data~\cite{OBH}, as shown, for example, by Fig.~\ref{fig_newPW}, where
$g^c_{\uparrow\uparrow}$ and $S^c_{\uparrow\downarrow}$ for $r_s=4$
obtained from our scheme~\cite{noi}, the revised PW 
model~\cite{comment} and QMC simulations~\cite{OBH} are reported.
We have checked that,
as $r_s$ increases, the discrepancy between revised PW and QMC data
becomes more pronounced. The reason of such increasing discrepancy
with descreasing density can be explained as follows.  
The revised
$\uparrow\uparrow$ PW function for the unpolarized gas is built up
by rescaling the pair--correlation function of the fully polarized gas
in such a way that the exact $k\to 0$ limit of $S^c_{\uparrow\uparrow}$
is fulfilled.
As a result, the parallel--spin contribution to correlation tends to be 
overestimated (and consequently the $\uparrow\downarrow$ part is
underestimated). Correlations are highly dominated
by antiparallel--spin interactions, where available 
(see e.g. Ref.~\cite{Pines}), and hence simple scaling arguments which connect
correlations in the fully--polarized gas
(where $\uparrow\downarrow$ interactions are totally
absent) and $\uparrow\uparrow$ 
correlations in the unpolarized gas 
(where $\uparrow\downarrow$ interactions are present and tend to
dominate the electronic correlations) will provide less and less reliable 
results at lower and lower densities, as the role
of Coulomb correlation with respect to exchange becomes more
and more important.
The same argument applies to the correlation energy, as we shall
see in the next Sec.~\ref{sec_energy}.
Notice that
the overestimate of the PW $\uparrow\uparrow$ pair--correlation function
makes the total (exchange $+$ correlation) 
parallel--spin pair--distribution function
be slightly negative near $r=0$ for densities $r_s\gtrsim 6$. 
Nonetheless, the new scaling law proposed
by Schmidt {\sl et al.}~\cite{comment} does much better than any previous
one.  
\section{Spin--resolved correlation energies and the high--density limit}
\label{sec_energy}
\begin{figure}
\begin{center}
\includegraphics[width=0.7\textwidth,height=0.3\textheight]{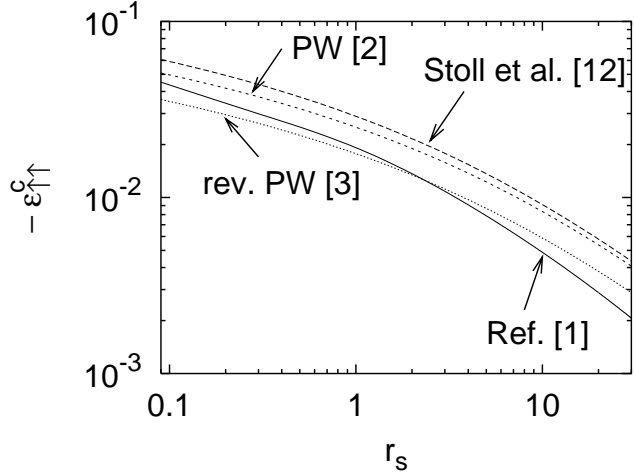}
\end{center}
\caption
{Parallel--spin contribution to the correlation energy
obtained from different scaling guesses compared to our 
interpolation scheme (solid line, Ref.~\cite{noi}). 
Units are Hartrees per electron.}
\label{fig_ecuu}
\end{figure}

{\it Correlation energy.\/}
The spin--resolved contributions to the correlation energy are
defined as
\begin{equation}
\epsilon^c_{\sigma_1\sigma_2}=\frac{q_F^2}{3\pi}\int_0^{r_s}
dr_s'\int_0^{\infty}d\rho\,\rho\,g_{\sigma_1\sigma_2}^c(\rho,r_s'),
\label{eq_ecspinres}
\end{equation}
and hence $\epsilon_c=\epsilon^c_{\uparrow\downarrow}+
\epsilon^c_{\uparrow\uparrow}$.
The corresponding exact high--density limit is recovered
by applying the same argument which yielded
the $k\to 0$ expansion of $S^c_{\sigma_1\sigma_2}$. In the
framework of RPA (see e.g. Ref.~\cite{Pines}), in fact, one obtains
for the unpolarized gas 
$\epsilon_{\uparrow\downarrow}^c=\epsilon_{\uparrow\uparrow}^c$ 
at any $r_s$.
Since in the $r_s\to 0$ limit RPA is exact, we have 
\begin{equation}
\lim_{r_s\to 0}\,\epsilon_{\uparrow\downarrow}^c=
\lim_{r_s\to 0}\,\epsilon_{\uparrow\uparrow}^c=
\tfrac{(1-\ln 2)}{2\pi^2}\,\ln r_s+O(r_s^0).
\label{eq_ecssHD}
\end{equation}
Beyond RPA (beyond orders $\ln r_s$) 
$\epsilon_{\uparrow\downarrow}^c$
and $\epsilon_{\uparrow\uparrow}^c$ are not equal because
of higher--order exchange terms which mainly lower the $\uparrow\uparrow$
correlation energy. The spin--resolved correlation energies
presented in Ref.~\cite{noi} are the best--to--date estimate
of $\epsilon_{\sigma_1\sigma_2}^c$, since they are obtained
by integrating the corresponding QMC pair--correlation 
functions~\cite{OBH} interpolated by our $g^c_{\sigma_1 \sigma_2}$
and $S^c_{\sigma_1 \sigma_2}$ models which also incorporate
the exact behavior at small
and large arguments. 
Moreover, our $\epsilon_{\sigma_1\sigma_2}^c$
fulfill the high--density limit of Eq.~(\ref{eq_ecssHD}). 
Previous estimates of $\epsilon_{\uparrow\uparrow}^c$ were obtained
by scaling the correlation energy of the fully polarized gas.
The most widely--used scaling laws are the Stoll~{\sl et~al.} 
\cite{Stoll} and the Perdew--Wang~\cite{PerdewWang}, which
is now available in its revised form given by Schmidt~{\sl et~al.}
\cite{comment}. In Fig.~\ref{fig_ecuu} we compare
our $\epsilon^c_{\uparrow\uparrow}$ with
these three schemes. It is apparent that the revised
PW gives the best result, even if it does not fulfill
the high--density limit of Eq.~(\ref{eq_ecssHD}).
As a consequence, the revised PW underestimates by $\sim 22\%$ 
the $\uparrow\uparrow$ correlation energy at $r_s=0$.
As expected from the corresponding $\uparrow\uparrow$
pair--correlation function (see Fig.~\ref{fig_newPW}),
at densities $r_s\gtrsim 2$
the revised PW overestimates $\epsilon^c_{\uparrow\uparrow}$ 
by an amount which increases with $r_s$
(e.g. 3\% at $r_s=3$, 21\% at $r_s=10$).
Again, we see the failure of simple scaling laws which try to
connect correlations in the fully polarized gas 
and $\uparrow\uparrow$ correlations in the unpolarized gas.

{\it High--density limit of the pair--correlation functions.}
It is also wortwhile to discuss some points about the link
between the high--density limit of the pair--correlation function
and the $r_s\to 0$ limit of Eq.~(\ref{eq_ecssHD}). 
Rassolov {\sl et al.}~\cite{ras} have recently computed the
$r_s\to 0$ limit of $g_c/r_s$, which turned out to be 
a well defined, $r_s$--independent, function.
For different reasons,
neither PW~\cite{PerdewWang,comment}, nor our pair--correlation
function~\cite{noi} fulfill this limit.
As pointed out
in Ref.~\cite{noi}, our simple functional forms do not reconcile
the known high--density limit of the pair--correlation functions
at zero interelectronic distance~\cite{g0HD} with Eq.~(\ref{eq_ecssHD}). 
In this respect, the functional form
used by PW~\cite{PerdewWang} is such that when the exact $k\to 0$ limit of
$S^c_{\sigma_1\sigma_2}$ is imposed to it, Eq.~(\ref{eq_ecssHD}) is
automatically violated. The PW spin--resolved
version, thus, suffers from a different problem than ours.
In summary, although major steps forward have been achieved
in this area by Refs.~\cite{noi,comment},
none of the existing models fulfills all the known
properties at $r_s\to 0$, and further improvements
are needed in this limit. 
 
\section{Conclusions}
In this work we have clarified the behavior of the 
spin--resolved pair--correlation
function and static structure factor of the unpolarized uniform electron
gas for small and large arguments. We have then compared
the two best--to--date models for these 
functions~\cite{noi,comment}, pointing
out their advantages and drawbacks, and we have discussed
the corresponding spin--resolved correlation energies. 
We found that the functions of Ref.~\cite{noi} provide
a better fit to the QMC data~\cite{OBH}, both in real
and reciprocal space, and provide the best--to--date
estimate of the spin--resolved correlation energies.
At very high density, however, even the forms proposed
by Refs.~\cite{noi,comment} are inadequate, and further
work is needed. It should be also kept in mind that in this
high--density limit the non--relativistic hamiltonian, on which
the whole theory is based, is no longer valid, and relativistic
corrections or a fully relativistic treatment wolud be
required~\cite{relativistic}. 

\begin{ack}
We are very grateful to P. Ballone for making available to us prior 
to publication the numerical results of Ref.~\cite{OBH}, and
to him and J.~P. Perdew for fruitful discussions. 
\end{ack}

\end{document}